\documentclass[article]{aastex631}

\shorttitle{In Violation of the Prime Directive}
\shortauthors{Fowler \& Murray-Clay}
\graphicspath{{./}{figures/}}

\begin{document}

\title{In Violation of the Prime Directive: Simulating detriments to Delta-Quadrant civilizations from the starship Voyager's impact on planetary rings}

\author[0000-0002-0726-9323]{J. Fowler}
\affiliation{University of California, Santa Cruz \\
1156 High St \\
Santa Cruz, CA 95064, USA}

\author[0000-0003-4549-0210-]{Ruth Murray-Clay}
\affiliation{UC Observatories \\
University of California, Santa Cruz \\
1156 High St \\
Santa Cruz, CA, 95060, USA}



\begin{abstract}

In the seven years that the starship Voyager spent in the Delta Quadrant, it used many questionable techniques to engage with alien civilizations and ultimately find its way home. From detailed studies of their logs and opening credits, we simulate Voyager's practice of orbiting a planet, to examine the effect on planetary rings. We outline a feasible planetary system and simulate the extent to which its rings would be disrupted. We find that Voyager's orbit could inflate the height of the rings in the vicinity of the spacecraft by a factor of 2, as well as increase the relative speeds of neighboring planetesimals within the rings. This increase in ring thickness has the potential to alter shadows on any moons of this planet, impacting ring-shadow based religions. Additionally, the acceleration of these planetesimals could rival their gravity, bucking any alien inhabitants and their tiny civilizations off of their planetesimal homeworlds. Finally, we posit that due to increased collisions amongst the planetesimals (which may harbor tiny intelligent life) the trajectory of these civilizations may be forever altered, violating the prime directive. 
\end{abstract}


\section{Introduction} \label{sec:intro}

The formation and evolution of planetary rings has been highly studied and modeled over the years \citep[for a review, see][]{goldreichtremaine, schmidt2009}.
Nbody simulations are a common tool to model the effect of rings given gravitational forces over time, including due to disruptions resulting from the changing orbits of moons. However, to date no published literature has tackled the impact of intelligently-driven foreign bodies on planetary dynamics and ring evolution. Here, we consider the problem of of how planetary rings may change after encountering the Starship Voyager. 

Star Trek: Voyager's \citep{voy} opening credits feature a collection of Voyager's exploits, presumably representative of their seven year journey in the Delta Quadrant of the Milky Way Galaxy. Much of the opening credits feature Voyager interacting with gas and dust, carving paths through nebulae and zooming around planetary rings. We aim to carefully treat a Voyager-planet interaction from the opening credits and examine impacts to the ring system. The Planetary Federation's Prime Directive states that no member of Star Fleet may interfere with the evolution and natural development of alien civilizations. \textbf{We examine to what extent simulations show that Voyager in this opening credit scenario could violate the Prime Directive.}

In Section \ref{sec:initial} we motivate our problem given a reconstruction of the opening credits and describe our initial assumptions to model a Voyager-planet interaction. In Section \ref{sec:sims} we describe numerical integrations of planetary ring dynamics conducted using the Nbody code \texttt{REBOUND} \citep{rebound, reboundias15}. In Section \ref{sec:discuss} we discuss potential impacts of ring-disruption and consider violations of the Prime Directive. Finally, in Section \ref{sec:conclusion} we make our conclusions and summarize recommendations for the protection of Delta-or-any-Quadrant aliens. 

\section{Initial conditions of a voyager-planet system} \label{sec:initial}

\begin{figure}
   \includegraphics[height=0.4\textheight] {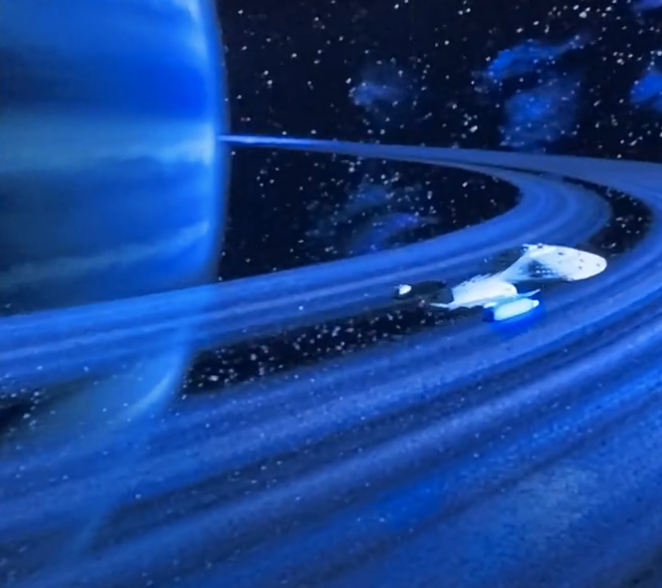}   
   \hspace{1em}
   \includegraphics[height=0.4\textheight] {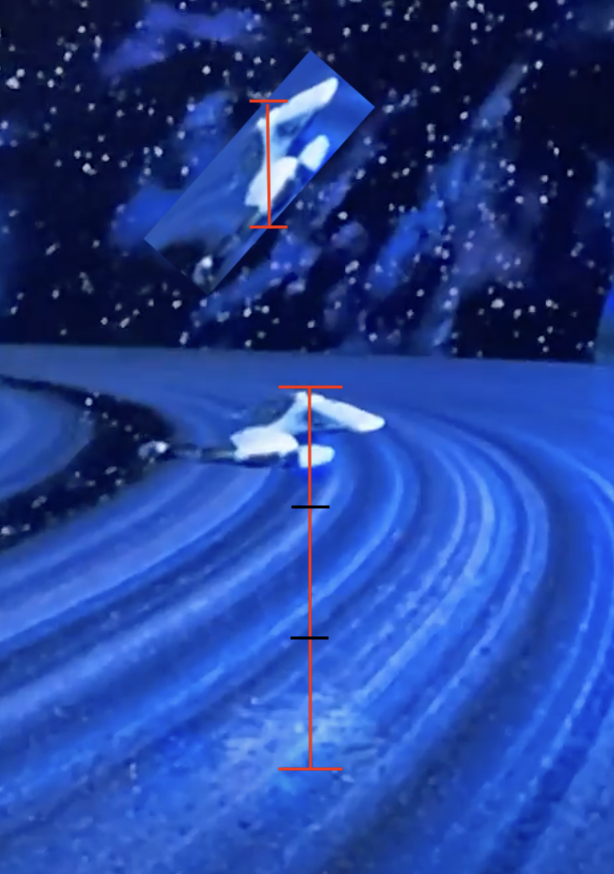}
\caption{Left: Video evidence from the opening credits allow us to estimate initial conditions. Right: Estimated height of Voyager above the rings of 3 Voyager lengths, based on reflection. Figure adapted from Star Trek: Voyager \citep{voy} 
\label{fig:voyager_plus_planet}}
\end{figure}

To estimate Voyager's impact on the rings of a given planet, we use a scene from the opening credits (shown in Figure \ref{fig:voyager_plus_planet}) to benchmark a typical Voyager-planet interaction. Assuming Voyager is the size of a middle school (we find this to be feasible based on the crew compliment it could support before the catastrophic run-ins of the first few episodes) we size Voyager at $\sim 200$ meters long.  Following standard theoretical procedure, we model Voyager as a sphere of radius $R_{\rm Voy} = 100m$. As illustrated using Voyager's reflection on the ring surface (Figure \ref{fig:voyager_plus_planet}, right panel), the spacecraft is approximately three Voyager-lenghths (6$R_{\rm Voy}$) above the ring surface. Using Voyager's size relative to the planet in Figure \ref{fig:voyager_plus_planet}, we rule out gas giant planets. The planet is blue in the opening credits, so we use a planet mass, $M_p$, and radius, $R_p$, equal to those of Neptune for our estimates. 

\begin{figure}
   \includegraphics[height=0.3\textheight] {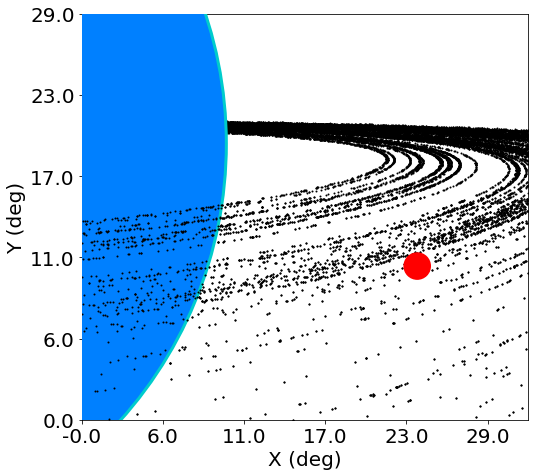}   
   \includegraphics[height=0.3\textheight]
   {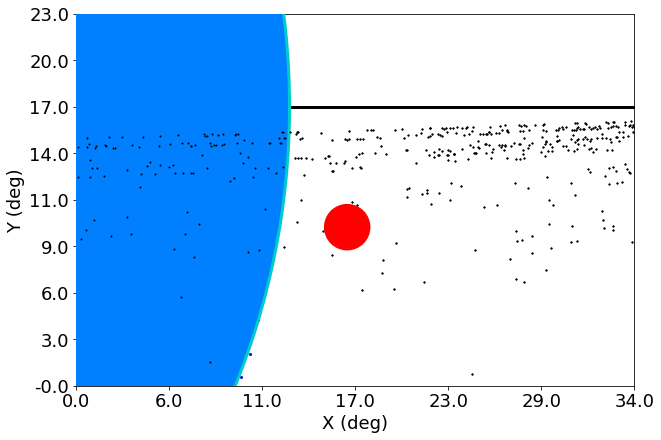}
\caption{Initial conditions of the simulations for the \texttt{REBOUND} setup. Left: Note that Voyager is 1000x too big in this plot for visual comparison with the original image. We expect that image enhancement \citep{enhance} was used to magnify the Voyager and the region around it -- leaving our use of Voyager as a measuring stick credible but allowing us to pick on other things we didn't like. Right: Here Voyager is a realistic size with respect to the planet, but the rings appear much thinner than in visual comparison. We expect this may be due to artificial projections of larger size, like that of the alien society in Catspaw \citep{tos}. 
\label{fig:initial_conditions}}
\end{figure}

The ring geometry in our simulations is illustrated in Figure \ref{fig:initial_conditions}.  For rings extending from 1.35--2.28$R_p$ (left panel), comparable to the rings of Saturn, we identify a viewing geometry comparable to that shown in Figure \ref{fig:voyager_plus_planet}.  However, given this geometry, Voyager itself must be enhanced in size by a factor of 1000 to have an appropriate size.  We expect that image enhancement \citep{enhance} was used to magnify Voyager in the opening credits sequence.  An alternative that does not require image enhancement is provided in the right-hand panel.  Here, Voyager is appropriately sized, but the rings are quite thin, extending from 1.3999 to 1.4001 planetary radii.  In both viewing geometries, Voyager orbits at a distance of 1.4$R_p$.  We prefer the wider rings due to a better visual match, but we note that our results for the dynamics of ring particles in the vicinity of Voyager are independent of which geometry we choose.

We assume Voyager's bulk composition resembles the density of Titanium at 4.5 g/cm$^3$. While Aluminum is more commonly used in spacecraft because it's light and sturdy, we assume that 4800-era Titanium alloys could provide the stability needed for warp-speed travel. With improved space mining and energy sourced from Dilithium crystals, we project an enriched Titanium alloy would be a suitable material. Voyager is not entirely metal, but likely contains $\sim$ two thirds empty space. Given this porosity, $\phi_{\rm Voy} = 0.66$, the spacecraft's density is  $\rho_{\rm Voy} = 1.53$ g/cm$^3$, and its mass $M_{\rm Voy} = 6.41\times10^{12} $ g $ =  1.07\times10^{-15}$ Earth masses. We note that the Hubble Space Telescope has a density of only 0.06 g/cm$^3$, but Voyager requires greater robustness because it travels faster, further, and has to survive more hull impacts.

While the size of particles in Saturn's rings can be as large as 1km, the typical size is 2-12 m (made mostly of water ice). \citet{hirata2022} argue that this maximum size is determined by thermal stresses and suggest that ring particles in the Neptunian region should also have similar maximum sizes.  We therefore assume that the largest ring particle sizes evident as the opening credits sweep through the rings is 1--10 m.  However, unlike Saturn's rings, which have a typical thickness of 10m (outside of radially-thin regions where perturbations by shepherding moons produce ``walls" that can reach 1km in height), these Delta-quadrant rings appear to be $\sim$30--100 particle-sizes thick. This puzzling difference may come from a distinct coefficient of restitution in Delta quadrant rings.  Indeed, \citep{schmidt2009} find that while rings modeled using a ``frosty" elasticity model are expected to be 10 particle-sizes thick, their preferred ``smooth" elasticity model is consistent with a 30 particle-size-thick ring that is marginally optically thick.

For a full-width ring thickness of 60m, our estimate that Voyager is 3 Voyager lengths above the rings (Fig \ref{fig:voyager_plus_planet}), puts the starship 10 ring thicknesses above the rings, consistent with the sweeping view provided in the opening credits. 
These starting assumptions are listed in Table \ref{tab:voy}. 

We simulate the impact of Voyager orbiting for 1 Earth-week on 2000 nearby planetary ring particles.  The ring particles (which may be 1--12 meters across at most) are small with respect to Voyager, so they are simulated as test particles  For computational efficiency, we restrict our simulated ring particles to orbits with semi-major axis randomly drawn from a uniform distribution between -2.5 and 2.5 $R_{\rm Voy}$ from Voyager's orbital semi-major axis at 1.4$R_p$.  The particles are assigned random inclinations from a uniform distribution 0 to $H/a_{\rm Voy}$, where $H = 30$m is the half-thickness of the rings and $a_{\rm Voy} = 1.4R_p$ is Voyager's orbital distance.  Longitudes of pericenter, $\omega$, and longitudes of ascending node, $\Omega$, are drawn uniformly between 0 and $2\pi$ radians, and ring particle true anomalies, $f$, are chosen such that the initial angular position of the particles ranges uniformly between $\pm 0.0002$ radians from the angular position of Voyager.  Voyager is initialized to be at its maximum vertical separation from the rings, with $\omega = 0$, $\Omega = \pi$, $f = \pi/2$, and inclination $H_{\rm Voy}/a_{\rm Voy}$ where $H_{\rm Voy}$ = 600 m.

\begin{deluxetable*}{llll}[h!]
\tablenum{1}
\tablecaption{Starting assumptions of the n-body simulation.\label{tab:voy}}
\tablewidth{0pt}
\tablehead{
\colhead{Parameter} & \colhead{Estimate} & \colhead{Justification} & \colhead{Notes} }
\startdata
planet mass & 1.02$\times10^{29} $g & Neptune mass & planet is blue in opening credits \\
average ring radius & 4.50$\times10^{9}$cm & 2 Neptune radii & Saturn-like rings \\
Voyager diameter & 200 m & $\sim$ middle school sized & bigger than a beach motel but smaller than a community college \\
Voyager height & 600 m & 3 Voyagers above the rings & see Fig \ref{fig:voyager_plus_planet} \\
Voyager total density & 1.53 g/cm$^3$ & 4.5 g/cm$^3$ at a porosity of 0.66 & Dilithium strengthened Titanium alloy  \\
\enddata
\end{deluxetable*}

\section{Simulated Voyager impacts}
\label{sec:sims}

We integrated the simulation forward using the orbital dynamics code \texttt{REBOUND} with the \texttt{ias15} integrator \citep{rebound, reboundias15}.   
We allow Voyager to orbit the planet for 7 days, a typical visit duration, and then examine impacts to the rings. Results are shown in Figures \ref{fig:azimuthal_gap} and \ref{fig:PUFF}.

\subsection{Ring puffing}

\begin{figure}[h!]
\begin{center}
   \includegraphics[width=\textwidth] {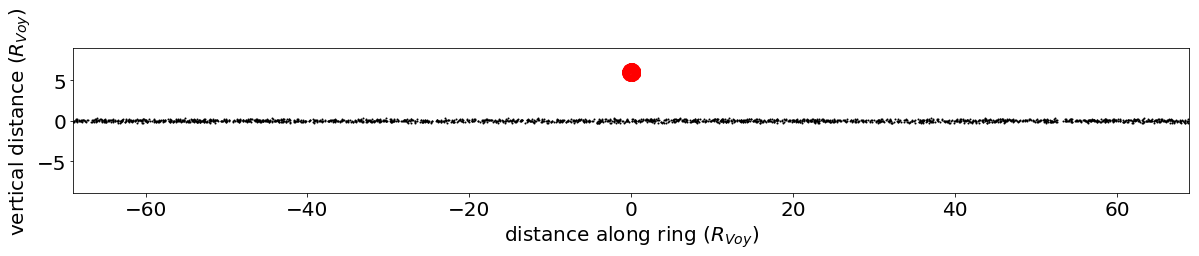}  
    \includegraphics[width=\textwidth] {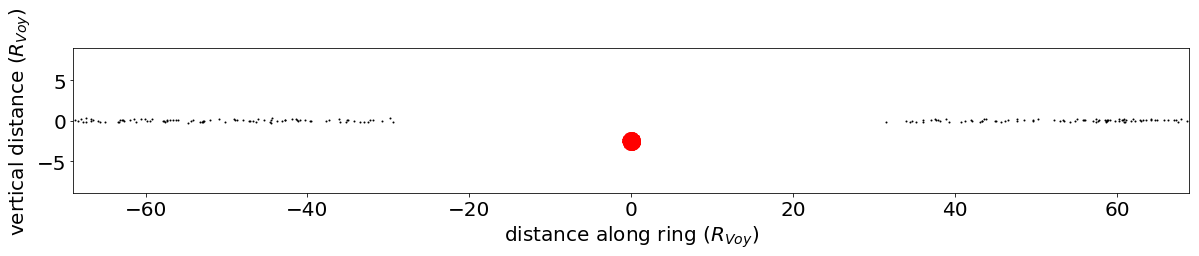}   
   \end{center}
\caption{Ring particle heights as a function of azimuthal distance along the ring (black points).  Initially (top), Voyager (red circle) lies above the ring.  After 7 days (bottom), Voyager's orbit has cleared an azimuthal gap extending a few tens of starship radii in both directions.  Voyager size is to scale. 
\label{fig:azimuthal_gap}}
\end{figure}

\begin{figure}[h!]
\begin{center}
   \includegraphics[width=0.48\textwidth] {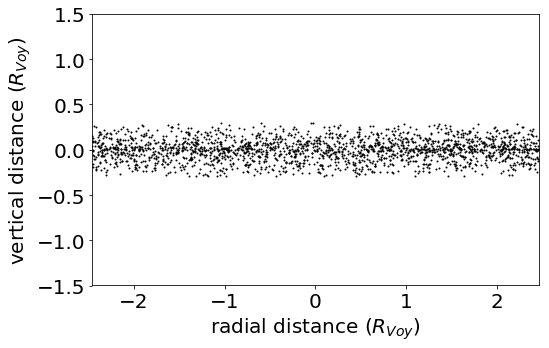}  
   \hspace{1em}
    \includegraphics[width=0.48\textwidth] {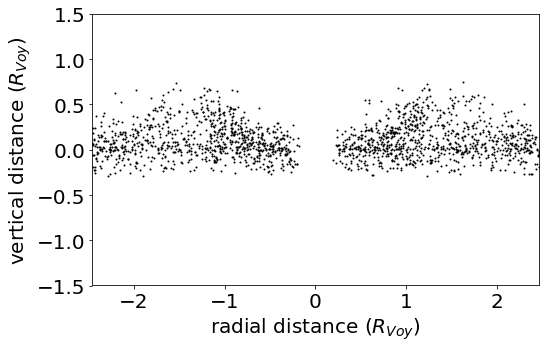}   
   \end{center}
\caption{Ring particle heights as a function of radial distance from the center of the planet.  Distances are measured with respect to the orbital separation of Voyager.  The smooth initial ring distribution (left) is vertically excited in the neighborhood of Voyager by approximately a factor of 2 (right) due to scattering by the starship.  A small radial gap is also evident at the orbital distance of Voyager. 
\label{fig:PUFF}}
\end{figure}

The major impact of the Voyager-planet interaction takes the form of increasing the thickness of the rings, or ring-puffing. After Voyagers 7-day orbit, the starship carves a small gap in nearby particles, both azimuthally (Figure \ref{fig:azimuthal_gap}) and radially (Figure \ref{fig:PUFF}.  Just beyond the gap, we see the nearby ring thickness increase by a factor of 2, as shown in Figure \ref{fig:PUFF}.  

The ability of Voyager to scatter nearby ring particles may be understood using the impulse approximation.  The initial ``random velocities" of ring particles (velocities relative to Keplerian) are approximately
\begin{equation}\label{eqn:vrand}
v_{\rm rand} = H\Omega  \sim 1 \textrm{cm/s} \;\;,
\end{equation}
where $\Omega = (GM_p/a_{\rm Voy}^3)^{1/2}$ is the orbital angular velocity.  The vertical velocity kick induced by Voyager over one 4.27-hour orbital period is approximately
\begin{equation}
v_{\rm kick} \sim \frac{GM_{\rm Voy}}{H_{\rm Voy}v_{\rm rel}} \sim \frac{2GM_{\rm Voy}}{H_{\rm Voy}^2\Omega} \sim  0.6 \textrm{cm/s} \;\;,
\end{equation}
where we have taken $v_{\rm rel} \sim H_{\rm Voy}\Omega$.

\subsection{Increased velocity and collision rate of ring particles}\label{sec:accel}

The random  velocities of the particles will increase with ring height (as described in Equation (\ref{eqn:vrand}) ; with a 2x increase in ring height, we project the particle speed will increase by approximately a factor of two from its initial value of $\sim$1 cm/s. 
This increase in speed over 7 days will create a puff-induced acceleration of $\sim 10^{-5}$ cm s$^{-2}$.  We note that the acceleration of some particles likely occurs on less than a 7-day timescale, so this estimate is a minimum acceleration.

\begin{figure}[h!]
\begin{center}
   \includegraphics[height=0.3\textheight] {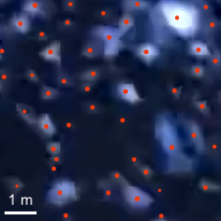}   
   \end{center}
\caption{Estimate of particle number density in the ring. We count 50 particles in this 5 meter (or 5 median-particle-size) box. 
\label{fig:particle_density}}
\end{figure}

In Figure \ref{fig:particle_density} we estimate a number density of particles in the rings. We assume the number density of particles stays constant across the puffing as particles ejected from the gap pile up at the gap edges, making up for the increased height.\footnote{This choice was entirely physically motivated and has nothing to do with the fact that we only have one screengrab of the opening credits from which to estimate number densities.} Using a median particle as 10 meters across we count 50 particles per 5x5m box (we count the box size as 5 median particles across), and take 30 meters (the ring height) as the depth, giving us a final number density of $n = 6.67\times10^{-8}$ cm$^{-3}$.  Using a geometric cross-section for collisions $\sigma$ and relative velocities, $v_{\rm rel}$ comparable to the random velocities of the disk particles, we calculate the collision rate: 
\begin{equation}
    \textrm{collision rate} = n\sigma v_{\rm rel} = 0.42 \textrm{ s$^{-1}$} 
\end{equation}
meaning we expect a collision of ring particles approximately every 2.5 seconds (where previously, it was only every 5 seconds). This results in $1.5\times10^{6}$ more collisions due to Voyager during its 7 day orbit. 


\section{Violations of the prime directive} \label{sec:discuss}

Seeing the notable effects that Voyager has on this ring system, we fear Voyager is likely to have a major impact on any life that exists within this ring system. Below we outline some cases in which the puffing of the rings, the increased velocity of the particles, and the increased collisions of the particles could violate the Prime Directive. 

\subsection{Creation of a super-space virus}

As evinced by numerous existing studies on space viruses \citep{tos, nextgen, ds9, voy}, one of the greatest dangers to space travel are viruses that have ranged from intoxication so powerful it leads to death (e.g., PSI 2000 of \cite{tos}), viruses that have disrupted communication (e.g., Aphasia Virus of \cite{ds9}), and viruses so large you can see them (e.g., Macroscopic virus of \cite{voy}). With increased collision of ring particles, foreign matter that would have otherwise remained dormant will be more likely to interact with other matter -- laying the groundwork for a super-space virus built from many ring-particle's worth of viral space matter. This could not only impact the crew of Voyager and other travelers to the region, but bring disease to any aliens that lived on moons or on the small ring particles. 

\begin{figure}[h!]
   \includegraphics[height=0.33\textheight] {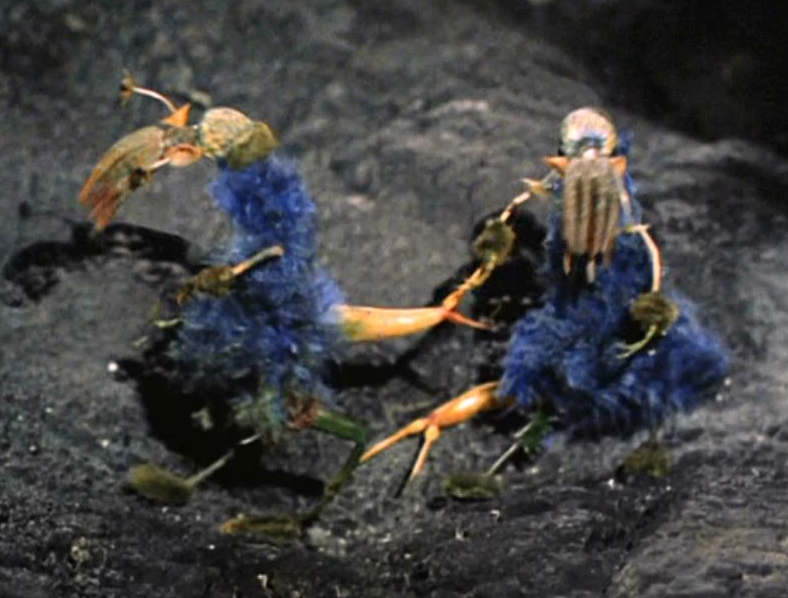}   
   \hspace{1em}
   \includegraphics[height=0.33\textheight] {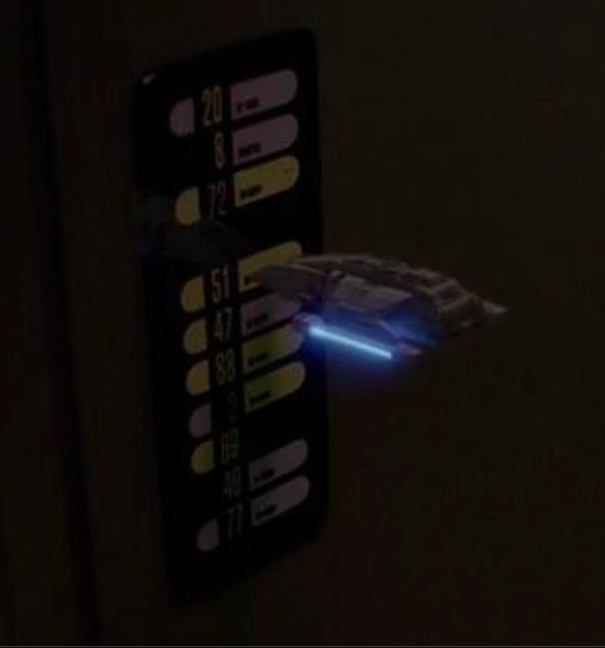}
\caption{Left: Morphology of tiny aliens after their humanoid projection is undone Figure adapted from Star Trek: The Original Series \citep{tos}. Right: Humanoids and the Defiant shrunken down to tiny size. Figure adapted from Star Trek: Deep Space Nine \citep{ds9}. 
\label{fig:tiny-aliens}}
\end{figure}


\subsection{Disruptions of tiny intelligent life living on ring particles}

Given previous surveys of alien civilizations (e.g., Catspaw aliens in \cite{tos}), as well as the feasibility of making humanoid-size aliens tiny (e.g., circumstance of One Little Ship in \cite{ds9}) it is not unlikely that the small planitesimals could harbor tiny intelligent life, see Figure \ref{fig:tiny-aliens} for examples. 

Creatures living on planetesimals and ring particles would have extremely low surface gravity; for the case of a 1 meter diameter particle with a density of 1 $g/cm^3$, their gravity would be $g_{p} = 1.4\times10^{-5}$ cm/s$^2$. This means that any of these creatures would be particularly vulnerable to acceleration of their planetesimals by increased particle velocity. The minimum Voyager-induced acceleration of $10^{-5}$ cm/s$^2$ estimated in Section \ref{sec:accel} which is comparable to the planetesimal gravity. 

Furthermore, due to the aforementioned increased collisions, the likelihood of two alien species who otherwise may not interact being thrust into each other increases greatly -- forever altering the course of their civilization. This could be notably traumatic for tiny aliens species, as one species could be \textbf{especially} tiny compared to another that was only somewhat tiny. 

\subsection{Interruptions to speculative shadow-based religions on ring-embedded moons}

If a moon were embedded in this planet's rings, the rings would cast a prominent band of shadow on the moon. We expect that for intelligent forms of life this shadow would inspire religious worship or festivals, and be closely monitored for seasonal variability. A two-times increase in the size of this shadow would certainly be noticeable. 
Saturn's rings contain abundant small gaps carved by moons, and similar gaps are clearly evident in Figure \ref{fig:voyager_plus_planet}.  Consider, for example, a moon located in the middle of a 30km-wide gap (similar to Saturn's Russell or Jeffreys gap).  
We estimate the shadow cast by the 60m-thick ring would take up $\sim 0.1$ degrees in the sky. 
However if Voyager is orbiting near the gap edge, then for the duration of the puffed rings, that would increase to $\sim 0.2$ degrees, nearly half the size of our sun in the sky. This changing ring shadow could be comparable to the Ferengi False Profits \citep{voy}, or the Earth Battle of Hayls \citep{herodotus}.

\section{Conclusions} \label{sec:conclusion}

In conclusion, we predict that Voyager could irreparably alter the course of many civilizations with its habit of orbiting planets. By simulating Voygaer as a middle-school sized object for 7 days in orbit around an opening-credits-like planetary system, we find that the system would undergo a ring puff that increases both the ring height and the random velocities of ring particles by up to a factor of 2 in the vicinity of the starship. The increased collision rate of the particles in this system could lead to a greater confluence of space material, potentially creating a super space virus. In terms of intelligent life, we predict that any tiny intelligent creatures on planetestimals within the rings would suddenly be thrust into each other, artificially disrupting the natural flow of their society. For smaller planetesimals, the puff-induced acceleration of these particles will be comparable to if not greater than their gravity, with high potential of bucking civilizations clear off their home-particle. These all represent clear violations of the Prime Directive, and should be addressed by future Star Fleet regulations. We recommend clear policies that avoid orbiting ringed planets with any signs of life---intelligent or otherwise. 


\begin{acknowledgments}
Thanks to the visionary mind of Gene Roddenberry for creating Star Trek, Arcelia Hermosillo Ruiz for her assistance watching Star Trek and poking fun of the Voyager opening credits, and Anne Datillo and Alexandra Mannings for their helpful consideration of the practicalities of a seasonal ring-shadow based religion. 
\end{acknowledgments}

\newpage

\software{\texttt{astropy} \citep{2018AJ....156..123A},  
          \texttt{matplotlib}, \citep{matplotlib},
          \texttt{numpy} \citep{2013RMxAA..49..137F}, 
          \texttt{REBOUND} \citep{rebound},
          }







\bibliography{voyager}{}
\bibliographystyle{aasjournal}



\end{document}